\documentclass{iau} 
\usepackage{graphicx}
\title[Surface convection in a 3D kinematic Dynamo Model] %% give here short title %%
{ A Three- dimensional Babcock-Leighton Solar Dynamo Model with Non-axisymmetric Convective Flows.}

\author[Hazra \& Miesch]   %% give here short author list %%
{Gopal Hazra$^1$
 \and Mark S Miesch$^2$}

\affiliation{$^1$Dept. of Physics, Indian Institute of Science, Bangalore - 560012,
India \\ email: {\tt hgopal@iisc.ac.in} \\[\affilskip]
$^2$National Oceanic and Atmospheric Administration, Boulder -80305, Colorado, USA \\email: {\tt Miesch@ucar.edu}}

\pubyear{2018}
\volume{340}  %% insert here IAU Symposium No.
\jname{Long-Term Datasets for the Understanding of Solar and Stellar Magnetic Cycles}
\editors{Dipankar Banerjee, Jie Jiang, Kanya Kusano \& Sami Solanki}
\begin{document}

\maketitle

\begin{abstract}
The observed convective flows on the photosphere (e.g., supergranulation, granulation) play a key role in the Babcock-Leighton (BL) process to generate large scale polar fields from sunspots fields. In most surface flux transport (SFT) and BL dynamo models, the dispersal and migration of surface fields is modeled as an effective turbulent diffusion. We present the first kinematic 3D FT/BL model to explicitly incorporate realistic convective flows based on solar observations. The results obtained are generally in good agreement with the observed surface flux evolution and with non-convective models that have a turbulent diffusivity on the order of $3 \times 10^{12}$ cm$^2$ s$^{-1}$ (300 km$^2$ s$^{-1}$).  However, we find that the use of a turbulent diffusivity underestimates the dynamo efficiency, producing weaker mean fields and shorter cycle.
\keywords{Sun: interior, Sun: magnetic fields, activity, Sun: Photosphere}
\end{abstract}

%\section{Introduction}
In the last two decades Babcock-Leighton (BL) solar dynamo models have grown to become the most promising paradigms to explain the origin of solar magnetic cycle and its irregularities (Choudhuri, Sch\"ussler \& Dikpati 1995, Karak et al. 2014). But in these models, the fragmentation and dispersal of the BMRs are executed by the convective flows on the photosphere (e.g., supergranulation and granulation) which is  modeled as a simple random of walk process treating it as an effective turbulent diffusion (Leighton 1964). In this paper we use explicit three-dimensional (3D) convective flow fields for the first time in a Babcock-Leighton dynamo model of the solar cycle.  On the surface, these flow fields are identical to the simulated flow fields from empirical model of Hathaway (2012) who used the observed line of sight velocity in the photosphere with SOHO/MDI instrument. However, here we extrapolate these flows below the surface to create a 3D rendition of surface convection that is responsible for the magnetic flux transport in the upper CZ.  Furthermore, in this initial implementation, we use a time-independent snapshot of the convective flow instead of the evolving flow fields used by Upton \& Hathaway (2014). 

We achieve this through the use of the 3D kinematic Surface flux Transport And Babcock-LEighton (STABLE) solar dynamo model (Miesch \& Teweldebirhan 2016; Hazra, Choudhuri \& Miesch 2017).  
%STABLE is a 3D model that explicitly places BMRs on the surface in response to the dynamo-generated toroidal field near the base of the CZ, so the BL process can operate more realistically than in previous 2D (axisymmetric) models that employ, for example, a non-local $\alpha$-effect. 
Since STABLE is 3D, the specified flow fields can include surface convection as well as differential rotation and meridional circulation.  Note that no existing global MHD convection simulation has sufficient resolution and scope to realistically capture surface convection on the scales of granulation to supergranulation.  So, our approach of specifying the convective velocity spectrum based on photospheric observations ensures that the surface convective motions are represented as realistically as is currently feasible.

\begin{figure}[t]
%\vspace*{-2.0 cm}
\label{fig1}
\begin{center}
\includegraphics[width=4.1in]{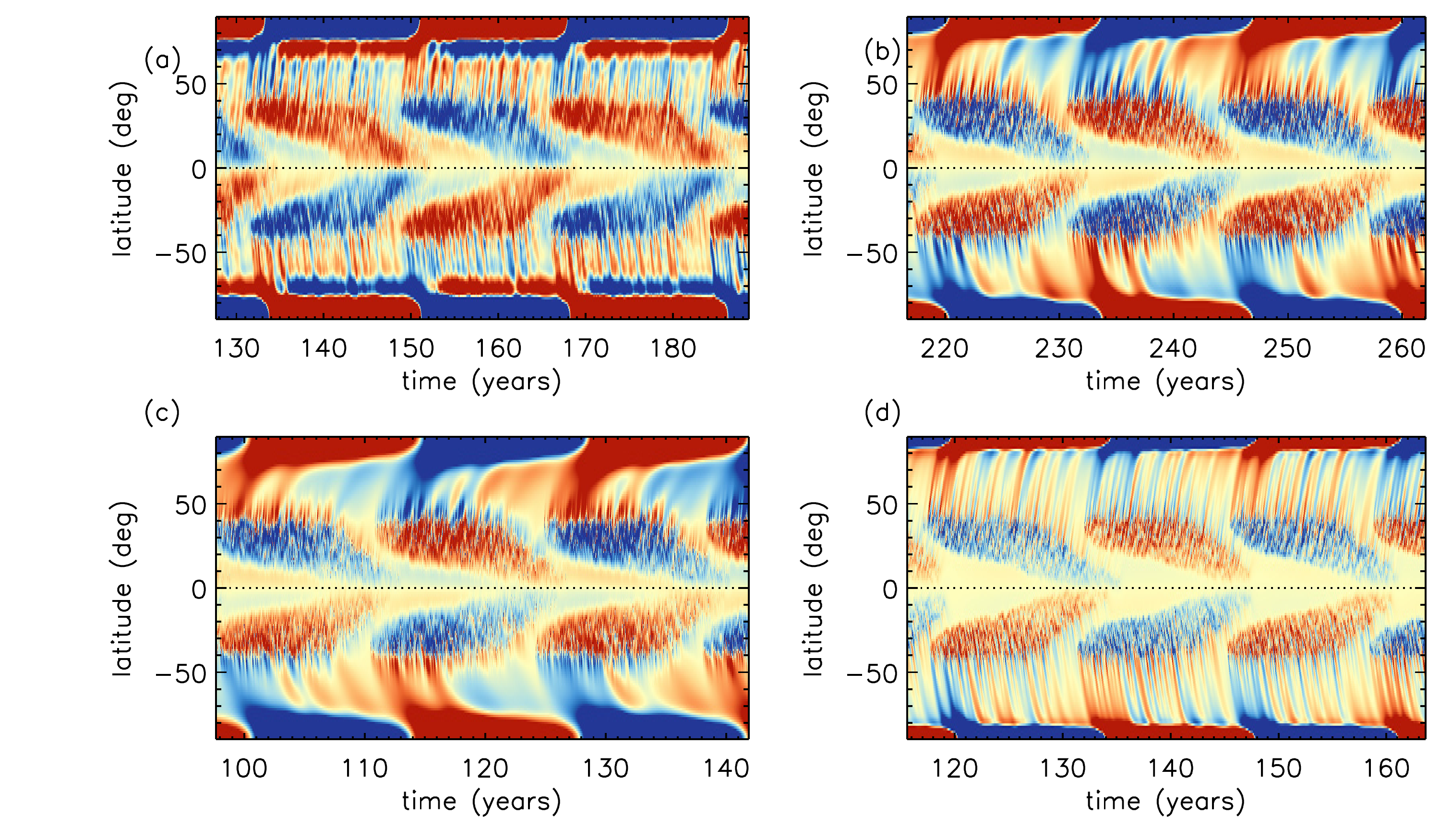} 
\caption{Time latitude plot of $\left<B_r\right>$ at $r=R$ for: (a) convective flows (b) $\eta_{top} = 3\times 10^{12}$ (c) $\eta_{top} = 10^{13}$ and (d) $\eta_{top} = 8\times 10^{11}$ cm$^2$ s$^{-1}$. Taken from Hazra \& Miesch (2018).}
\end{center}
\vspace*{-0.2 cm}
\end{figure}

Soon after implementing the convective flows, we realize that convective flows do operate as a small scale dynamo disrupting the large scale magnetic cycle. The only effective way we found to suppress the small-scale dynamo action is by making the convective flows to act on the vertical radial magnetic fields only. Also, we have used a background diffusivity ~$5 \times 10^{11}$ cm$^2$ s$^{-1}$ to maintain dipolar parity. This background diffusivity helps to suppress the small-scale dynamo action as well.

The butterfly diagram for the case with the convective flows is shown in Fig~1(a). This is comparable with the observed evolution of the radial fields on the surface of the Sun but the presence of mixed polarity near the pole has no counterpart in the observation. Such behavior may be attributed to the tendency for the convective motions to disperse and transport BMR fields without dissipating them. As a result, both polarities are transported poleward and concentrated into strong, alternating bands. 
%Usually with turbulent diffusion models, BMRs emerge and the opposite polarities partially cancel each other as they disperse.  By contrast, in Case of convective flows, the fields disperse but cancellation is less efficient as a result of the smaller ohmic diffusion.  

We also want to check whether the convective transport accurately parameterized by a turbulent diffusion or is it fundamentally non-diffusive? If the former then what value of surface diffusivity ($\eta_{top}$) is optimal? To do that, we have compared the surface butterfly diagram in case of convective flows to several diffusive cases with relatively high and low values of $\eta_{top}$, ranging from $1 \times 10^{13}$ cm$^2$ s$^{-1}$ to $8 \times 10^{11}$ cm$^2$ s$^{-1}$ as shown in Fig~1. Qualitatively, it bears the greatest resemblance to Case with turbulent diffusion $3 \times 10^{12}$ cm$^2$ s$^{-1}$ (Fig~1(b)). This conclusion is based on the width of the poleward-migrating streams (though note the different ranges for the time axes), the structure and width of the polar flux concentrations, the relative strength of the polar and low-latitude fields, and the location of the active latitudes. But based on the study of dynamo efficiency, migration speed of the poleward flux and mean field emf analysis, we find that modeling convective flux transport as a turbulent diffusion over-estimates the ohmic dissipation producing weaker mean fields and shorter cycle.


\vspace*{-0.5 cm}
\begin{thebibliography}{}
\bibitem[{{Choudhuri} {et~al.}(1995){Choudhuri}, {Sch\"ussler}, \&
  {Dikpati}}]{CSD95}
{Choudhuri}, A.~R., {Sch\"ussler}, M., \& {Dikpati}, M. 1995, \textit{A \& A}, 303, L29

%\bibitem[{Hathaway {et~al.}(2000)Hathaway, Beck, Bogart, Bachmann, Khatri,
%  Petitto, Han, \& Raymond}]{hatha00}
%Hathaway, D.~H., Beck, J.~G., Bogart, R.~S., Bachmann, K.~T., Khatri, G.,
%  Petitto, J.~M., Han, S., \& Raymond, J. 2000, Solar Physics, 193, 299
  
\bibitem[{{Hathaway}(2012)}]{Hathaway12a}
Hathaway, D. H. 2012, \textit{ApJL}, 749, L13

\bibitem[{{Hazra} {et~al.}(2017){Hazra}, {Choudhuri}, \& {Miesch}}]{HCM17}
{Hazra}, G., {Choudhuri}, A.~R., \& {Miesch}, M.~S. 2017, \textit{ApJ}, 835, 39

\bibitem[{{Hazra} \& {Miesch}}]{HM18}
{Hazra}, G. \& {Miesch}, M.~S. 2018, \textit{ApJ}, 864, 110

\bibitem[{{Karak} {et~al.}(2014){Karak}, {Jiang}, {Miesch}, {Charbonneau}, \&
  {Choudhuri}}]{Karakreview14}
{Karak}, B.~B., {Jiang}, J., {Miesch}, M.~S., {Charbonneau}, P., \&
  {Choudhuri}, A.~R. 2014, \textit{space. sci. rev.}, 186, 561

\bibitem[{{Leighton}(1964)}]{Leighton64}
{Leighton}, R.~B. 1964, \textit{ApJ}, 140, 1547

%\bibitem[{{Miesch} \& {Dikpati}(2014)}]{MD14}
%{Miesch}, M.~S. \& {Dikpati}, M. 2014, \textit{ApJL}, 785, L8

\bibitem[{Miesch \& Teweldebirhan(2016)}]{MT16}
Miesch, M.~S. \& Teweldebirhan, K. 2016, \textit{Adv.\ Space Res.}, 58, 1571

\bibitem[{Upton \& Hathaway(2014)}]{upton14b}
Upton, L. \& Hathaway, D. 2014, \textit{ApJ}, 792, 142 (7 pp.)


%\bibitem[{Upton \& Hathaway(2014)b}]{upton14a}
%Upton, L. \& Hathaway, D. 2014b, \textit{ApJ}, 780, 5 (8 pp.)

\end{thebibliography}
\end{document}